\PassOptionsToPackage{table}{xcolor}
\documentclass[11pt]{article}

\usepackage[final]{acl}
\usepackage{times}
\usepackage{latexsym}
\usepackage[T1]{fontenc}
\usepackage[utf8]{inputenc}
\usepackage{microtype}
\usepackage{inconsolata}
\usepackage{graphicx}
\usepackage{tabularx}
\usepackage{amsmath}
\usepackage{booktabs}
\usepackage{multirow}
\usepackage{geometry}
\usepackage{siunitx}
\usepackage{xcolor}
\usepackage{dashrule}
\usepackage{verbatim}
\usepackage{tcolorbox}
\usepackage{fancyvrb}
\usepackage{fvextra}
\tcbuselibrary{breakable}
\usepackage{paralist}
\usepackage{dsfont}
\usepackage{balance}
\usepackage{etoolbox}
\makeatletter
\patchcmd{\paragraph}{\normalsize\bfseries}{\if@nobreak\else\needspace{4\baselineskip}\fi\normalsize\bfseries}{}{}
\patchcmd{\subsection}{\normalsize\bfseries\raggedright}{\needspace{4\baselineskip}\normalsize\bfseries\raggedright}{}{}
\patchcmd{\section}{\large\bfseries\raggedright}{\needspace{3\baselineskip}\large\bfseries\raggedright}{}{}
\makeatother
\usepackage{needspace}
\usepackage{amsmath, amssymb, amsthm}
\theoremstyle{plain}

\newcommand{\rsi}{MedForge-RSI}

\newcommand{\todo}[1]{\textcolor{red}{\textbf{[#1]}}}

\title{MedForge-RSI: Recursive Self-Improvement\\for Medical Deepfake Detection}

\author{
 \textbf{Zhihui Chen\textsuperscript{1}},
 \textbf{Mengling Feng\textsuperscript{1}\thanks{Corresponding author}}
\\
\\
\textsuperscript{1}National University of Singapore\\
 \texttt{zhihui.chen@u.nus.edu, ephfm@nus.edu.sg}
}

\begin{document}
\maketitle

\begin{abstract}
Text-guided editors generate medical deepfakes at high fidelity, and reasoning-based detectors address this threat with strong results: MedForge-Reasoner, an 8B vision--language model trained with supervised fine-tuning and reinforcement learning, reaches 99.2\% accuracy in its target distribution. Under deployment conditions, however, the frozen model wrongly rejects 40\% of authentic scans and falls to 77\% on unseen generators and 59\% under transmission distortion; moreover, the conventional remedy is costly, since every retraining run requires 50{,}000 supervised images and expert-written guidelines. We present \rsi{}, in which the deployed detector improves itself while its weights remain frozen: in each of 20 rounds it examines its verified errors, records experience entries, and writes its own image-analysis tools; every proposed change is admitted only after passing an acceptance test on an independent validation split. Across 49 registered configurations, this procedure raises the four-test-set average from 75.0\% to 84.4\% and, on a 4,000-image final evaluation, clean accuracy from 76.5\% to 87.9\% and accuracy under transmission distortion from 59.0\% to 70.5\%. The improvements concentrate on real-image recall, the error type that remains at 60\% after reinforcement learning. A controlled analysis of all 49 trajectories identifies which mechanisms replicate across seeds and which fail, isolates acceptance testing as the largest single component, and yields a failure taxonomy; we release complete trajectories including every rejected and rolled-back change.
\end{abstract}

\section{Introduction}
\label{sec:intro}

Medical images have become an attractive target for text-guided deepfake generation, as a short instruction suffices for an editor to implant a tumor that was never present or to erase a lesion that is present \cite{diffusion_editors_survey,chen2025med}. Editors such as FLUX.1, GPT-Image, Nano-Banana, and the medical-specific Med-Banana \cite{flux,gptimage,nanobanana,chen2026medbananalearningqualitycontrolledmedical} perform such manipulations on brain MRI, chest radiographs, and fundus photographs with few visible artifacts, and a deepfake that passes visual inspection can compromise diagnosis, treatment, or insurance decisions \cite{miccai-detect,DeepFakesinHealthcare}. MedForge \cite{chen-etal-2026-medforge} addresses this threat at training time: its 8B vision--language model is fine-tuned on 50{,}000 forgery-aware reasoning chains and then optimized with group-sequence policy optimization (GSPO), reaching 99.2\% accuracy in the target distribution while localizing and explaining each forgery.

Once the same frozen model is deployed, three weaknesses emerge that retraining does not resolve. First, its errors are asymmetric: fake-image recall is 99\%, but real-image recall is only 60\%; consequently, 40\% of authentic scans are falsely reported as forged, the failure mode most damaging to clinical trust \cite{sida,fakeshield}. Second, accuracy declines under distribution shift, falling to 77\% on a generator family excluded from all evaluation pools and to 59\% on images distorted by the resizing, printing, and recompression that occur when medical images circulate in practice. Third, the conventional remedy scales poorly: each retraining run consumes 50{,}000 supervised images, a reinforcement learning stage, and a medically authored guideline, and must be repeated for every hospital, scanner, and newly released editor.

This paper investigates whether the deployed detector can itself improve at deployment time while its weights remain frozen. Research on recursive self-improvement (RSI) provides a suitable framework. A recent survey defines the harness around a model, namely its memory, tools, and orchestration code, and characterizes harness self-modification as \textquotedblleft the most concrete form of~`the agent rewrites itself.'\textquotedblright\ \cite{chen2026rsisurvey}; systems in this family report gains of 8 to 30 points on coding and video benchmarks \cite{zhang2026dgm,xu2026videoharness}. The same literature records characteristic failure modes, ranging from self-confirming evaluation to objective hacking \cite{chen2026rsisurvey,duan2026lastai}, and observes that the strength of demonstrated improvement scales with the reliability of the verification signal.

Our setting nevertheless remains outside this literature in two respects. No RSI system of this kind has been studied in a safety-critical perception domain: existing self-evolving systems in medicine target clinical reasoning and treatment workflows \cite{li2025agenthospital,duan2026lastai}, leaving perception unaddressed, even though deepfake detection supplies exactly the strong verification signal that this literature identifies as the binding resource, namely a ground-truth label for every image. In addition, the improved artifact in existing systems is typically a single program, workflow, or library \cite{zhang2026dgm,xu2026videoharness}; the authors of the closest system state this limitation themselves, noting that it constructs per-video contexts rather than memory that persists across questions \cite{xu2026videoharness}. To our knowledge, the combination we require, in which experience retention and tool creation serve as parallel knowledge stores under a safety architecture for admitting them, remains unexplored.

\begin{figure*}[t]
\centering
\includegraphics[width=\textwidth]{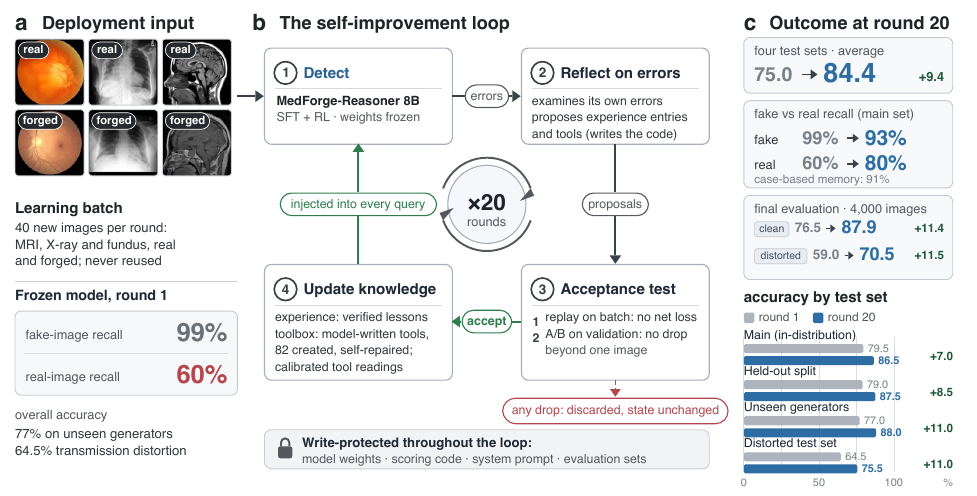}
\caption{\textbf{The \rsi{} loop.} \textbf{a}, At deployment the frozen MedForge-Reasoner processes a fresh learning batch of 40 labeled images in every round, and its errors are asymmetric: fake-image recall is 99\% but real-image recall only 60\%. \textbf{b}, Reflection examines these errors and proposes changes to two knowledge stores, structured experience entries and sandboxed tools written by the model itself. Every proposal must pass a two-stage acceptance test, consisting of replay on the learning batch and an A/B comparison on an independent validation split, and a proposal that reduces accuracy on either check is discarded. The model weights, the scoring code, the official system prompt, and the evaluation sets are write-protected throughout. \textbf{c}, Over 20 rounds, without any parameter update, the strongest of the 49 registered configurations raises the four-test-set average from 75.0\% to 84.4\% and real-image recall from 60\% to 80\%, or to 91\% with case-based memory, and on a separate final evaluation of 4,000 held-out images clean accuracy rises from 76.5\% to 87.9\% and accuracy under transmission distortion from 59.0\% to 70.5\%.}
\label{fig:system}
\end{figure*}

We therefore build \rsi{} around the frozen detector (Figure~\ref{fig:system}). The system maintains two knowledge stores. The experience memory holds structured entries that record an imaging context, a visual cue, and a caution, and usage statistics determine when an entry is retired. The toolbox contains tools written by the detector itself, which it repairs, retires, and re-enables autonomously. In every round the detector reflects on its verified errors and proposes changes to either store, and every proposal must pass a two-stage acceptance test consisting of replay on the learning batch followed by an A/B comparison on an independent validation split that is disjoint from every test set. The model weights, the scoring code, the official system prompt, and all evaluation sets remain outside the loop's write access throughout. For detection safety this is the decisive property: the detector remains effective against the deepfakes that successive editors produce, under an admission mechanism that ensures every change is verified, attributable, and reversible, rather than exchanging the frozen model's progressive obsolescence for an unverified adaptive component.

We evaluate 49 registered configurations, each running 20 rounds over a fixed learning stream, on four 200-image test sets and on a final evaluation of 4,000 images held out from all training images. Self-improvement under the acceptance test raises the four-test-set average from 75.0\% to 84.4\% and final-evaluation clean accuracy from 76.5\% to 87.9\%, with real-image recall rising from 60\% to 80\% (91\% with case-based memory) and unseen-generator accuracy from 77\% to 89.5\%. The remainder of the paper is organized around four findings from these runs: acceptance testing contributes more than reflection quality does; self-written tools must be judged on calibrated measurements rather than on their textual output; the best result of a single run often fails to replicate across learning streams; and the verifier itself can fail silently and therefore requires health checks, as the loop itself requires acceptance tests.

Our contributions are fourfold:
\begin{compactitem}
\item \textbf{System.} \rsi{}, to our knowledge the first systematic study of recursive self-improvement in a safety-critical perception domain, built on a published SFT+RL detector with no additional training compute.
\item \textbf{Two-store architecture.} Experience entries combined with self-written tools under independent-validation acceptance testing, improving the frozen detector by 9.4 points on four test sets and 11 points on the final evaluation.
\item \textbf{Controlled study.} 49 registered configurations isolating mechanisms that replicate across seeds (calibration, ensemble reflection, two-sided reflection, case-based memory) and mechanisms that fail (tournament selection, admission budgets, wholesale policy rewrites).
\item \textbf{Failure taxonomy.} Four measured failure modes with safeguards, with release of complete trajectories including every rejected and rolled-back change.
\end{compactitem}

\needspace{8\baselineskip}
\section{Related Work}
\label{sec:related}

\paragraph{Medical deepfake detection.}
Early medical forgery detectors are black-box classifiers trained on pairs of authentic and manipulated images \cite{miccai-detect,meddetect-from-miccai,DeepFakesinHealthcare}, and the deepfake threat model now spans copy-move tampering \cite{ctgan,forgerymedical}, text-guided diffusion editing \cite{diffusion_editors_survey,chen2025med}, and editors trained directly on quality-controlled medical editing trajectories \cite{chen2026medbananalearningqualitycontrolledmedical}. A second line of work uses multimodal large language models to explain forgeries \cite{sida,fakeshield,fakevlm,holmes}, but the explanations are post hoc and frequently hallucinated. MedForge \cite{chen-etal-2026-medforge} reframes detection as localize-then-analyze reasoning and reaches 99.2\% accuracy in the target distribution. All are static: after training, neither weights nor decision procedure changes. Medical self-evolution targets clinical reasoning and treatment workflow, as in Agent Hospital's MedAgent-Zero clinicians \cite{li2025agenthospital}; perception is unstudied. \rsi{} takes the strongest of these systems as a frozen base and examines what can be gained at deployment without retraining.

\paragraph{Recursive self-improvement at deployment.}
Within-episode self-refinement \cite{madaan2023selfrefine,shinn2023reflexion} improves one output without persistent change, whereas the persistence gradient identified in that survey motivates improving the surrounding software instead \cite{chen2026rsisurvey}. Persistent skill libraries originate with Voyager \cite{wang2023voyager}, and search over the agent's own components extends from agent-design search \cite{hu2024adas,zecheng2025aflow} and self-referential code modification \cite{yin2025godelagent}, descending from the original G\"odel machine \cite{schmidhuber2003godel}, to the Darwin G\"odel Machine \cite{zhang2026dgm}, which evolves the agent's own codebase. Closest to our setting, VideoHarness-RSI \cite{xu2026videoharness} improves context-construction programs around a frozen 8B vision--language model for long-video understanding, gaining 8.2 points under strict promote-only-if-strictly-better rules; its authors state the limitation that these programs construct per-video contexts rather than memory persisting across questions. In perception, ActiveScope \cite{wang2026activescope} self-corrects by actively re-inspecting visual regions within one episode. \rsi{} differs in the domain, safety-critical medical perception with label-verifiable errors, in improving two parallel stores rather than one program, and in independent-validation acceptance testing.

\paragraph{Experience memory and program search.}
A parallel line of work persists experience rather than skills: ExpeL \cite{zhao2024expel} distills insights from success and failure trajectories, ReasoningBank \cite{ouyang2026reasoningbank} turns self-judged trajectories into retrievable strategies, ACE \cite{zhang2025ace} curates the agent's context as an evolving ``playbook'', and ALMA \cite{xiong2026alma} meta-learns the memory design; none updates weights. Our experience entries add two restrictions: each entry is individually admitted or retired by the acceptance test, and usage statistics rather than the model's own judgment determine persistence. FunSearch \cite{romeraparedes2024funsearch} and AlphaEvolve \cite{novikov2025alphaevolve} evolve external artifacts rather than the agent's own mechanisms, and the survey accordingly classifies this line of work under automated research \cite{chen2026rsisurvey}; our tool-creation channel descends from Voyager's self-created skills and Agent0's tool-integrated self-play \cite{xia2026agent0}, and since SkillsBench measures no gain from LLM-authored skills \cite{gautam2026skillaxe}, each tool is judged by calibrated readings and the acceptance test.

\paragraph{Training-time self-improvement.}
A parallel line of work enables models to generate their own training signal, from self-generated rationales \cite{zelikman2022star} and self-play fine-tuning \cite{chen2024spin,yuan2024selfrewarding} to zero-data self-play verified by code execution \cite{zhao2025absolutezero,yue2026drzero}. These methods update weights and require training infrastructure that is unavailable at deployment in clinical settings. The resource they depend on, cheap ground-truth-grounded feedback, is exactly the one we exploit without training.

\paragraph{Surveys, verification, and safety.}
\citet{chen2026rsisurvey} classify 1,250 papers along what is improved (deployment-time, training-time, self-evaluation, automated research) and who verifies (human-in-the-loop, human-on-the-loop, closed loop), and identify four failure modes: self-confirming loops, model collapse \cite{shumailov2024modelcollapse}, diversity collapse, and frame lock-in. Audits of automated-research agents \cite{rank2026posttrainbench} document reward hacking without inducement; our sandbox and hash-committed evaluation sets are designed to prevent it. Verifier reliability is likewise domain-dependent: zero-shot detectors of machine-generated text degrade under domain shift, and domain-aware scoring recovers much of the loss \cite{chen-etal-2025-divscore}. \rsi{} occupies the deployment-time $\times$ human-on-the-loop cell, which already contains the coding and video systems above, and supplies its safety-critical perception instance: the loop proposes and validates autonomously while the validation split, scoring code, and model weights remain fixed, following the protected-core design of \citet{shi2026aevolve} and the bounded, human-auditable rule policies that \citet{shi2026sharp} evolve for high-risk domains. Our failure taxonomy (\S\ref{sec:analysis}) instantiates these warnings with measured instances.

\section{\rsi{}: Self-Improvement with Acceptance Testing}
\label{sec:method}

\subsection{Problem Formulation}
\label{sec:formulation}

This section formalizes the loop's state and its acceptance test (\S\ref{sec:formulation}), then details the two knowledge stores (\S\ref{sec:channels}), the reflector (\S\ref{sec:reflection}), and the protected components (\S\ref{sec:protected}).

Let $M$ be a deployed detector, MedForge-Reasoner, an 8B vision--language model produced by SFT and GSPO \cite{chen-etal-2026-medforge}, whose weights are permanently frozen. Following the formalism of \citet{chen2026rsisurvey}, everything that turns $M$ into a working system is explicit state that the loop may modify under rules we define: a memory of experience entries $E$, a toolbox $T$ of executable image-analysis tools, and a directive $d$ of detection reminders injected into the user turn. A prediction is $y = M(x \mid \mathrm{inj}(E, T, d))$, where $\mathrm{inj}$ renders retrieved entries, tool readings, and the directive into the model's context; verdicts are parsed from a fixed conclusion line by scoring code that the loop cannot modify.

Each round $t$ presents a fresh learning batch $B_t$ of 40 labeled images, never reused across rounds, which the system classifies before observing the errors. A reflector $\rho$, the detector itself in our default configuration, examines the errors and proposes a finite set of changes $\Delta_t$, comprising new or retired experience entries together with tool actions of the form create, modify, enable, or disable, and the state updates to $S_t \oplus \Delta_t$ if an acceptance test $G(\Delta_t; S_t, B_t, V)$ approves, and remains at $S_t$ otherwise, where $V$ is a validation split disjoint from every test set (\S\ref{sec:protected}). The design target is bounded self-refinement in the sense of \citet{chen2026rsisurvey}: improvement measured against a fixed external evaluator, with every persistent change individually attributable, testable, and reversible. Test sets are used only for scoring and never enter $G$, reflection, or early stopping.

\subsection{Two Knowledge Stores}
\label{sec:channels}

\paragraph{Structured experience memory.}
Each entry is a typed record $(\textit{context}, \textit{cue}, \textit{caution})$ that specifies the imaging context in which it applies (modality, generator family, distortion type), the visual cue to look for, and the caution it encodes; at most five entries are rendered into any single query. One representative entry reads: in fundus images produced by inpainting editors, look for disc-boundary asymmetry with vessel override, and treat symmetric peripapillary changes as normal variants. Entry conditions carry an explicit fallback to a small set of general entries, so any period in which the library is not consulted is detected. Every entry also accumulates usage statistics, counting the injections that coincided with correct verdicts and those that coincided with incorrect ones, and entries whose statistics or test performance decline are retired by a fixed rule. A focus-rotation schedule cycles the reflector through diagnostic themes and presents the reflector with its own directive history, which prevents the single-theme repetition observed in early runs.

\paragraph{Autonomous toolbox.}
The toolbox is initialized with five human-written tools (error-level analysis, FFT spectrum, a JPEG block grid, a noise residual measure, and an external second-opinion tool) and is thereafter managed by the model. Proposals may enable, disable, modify, or create tools under a strict contract: each tool is a single self-contained file that uses only Pillow and NumPy, raises no exceptions, and returns deterministic output capped at 600 characters. The sandbox executes each candidate on probe images with a 15-second timeout, and failures are returned to the reflector, which repairs its own code (\S\ref{sec:tools} quantifies the repair cycles). Created tools return raw measurements and never verdicts, as the contract forbids conclusion labels and thereby prevents a tool from introducing a classification decision through its output text.

\paragraph{Tool-reading calibration.}
Tools written by the model prove to be unreliable instruments: a majority of the implementations in our study emitted output from which no directional reading could be parsed, and a quarter embedded hardcoded numbers in their return strings while computing a quantity unrelated to the reported value (\S\ref{sec:tools}). The system therefore maintains, for every tool, the empirical distribution of its readings on historical images whose ground truth is known, and translates each raw reading into directional evidence, namely its percentile within the fake-image and real-image reading distributions. A reading is rendered as a calibrated statement such as ``91st percentile of historical \textsc{fake} readings vs.\ 34th of \textsc{real}, leaning \textsc{fake},'' or as \textsc{no signal} when the two distributions overlap. This component is the most reliable positive mechanism in the study. It converts fabricated or uninformative numbers into calibrated evidence, and the percentile anchor does not drift when directive text is rewritten.

\subsection{Reflection: Who Writes the Proposals}
\label{sec:reflection}

In the default configuration the detector reflects on itself in a minimal question--answer session with the official system prompt removed, so that it is not anchored to its current decision policy. Reflection proceeds in two stages, a free-form diagnosis of the misclassified images followed by strictly parsed JSON proposals. Three reflector variants are isolated experimentally: an external reflector implemented with a strong instruction-tuned LLM; ensemble reflection, in which three independent reflections at different temperatures and perspectives are merged before testing; and two-sided reflection, which reports false alarms and missed forgeries in a combined list and requires every proposal to declare which direction it addresses and to present evidence that the opposite direction is unharmed. A fourth mechanism, case-based memory, avoids generation risk altogether, since verified error cases are archived, up to 30 in our configuration, and retrieved at detection time as factual precedents.

\subsection{Acceptance Testing and Protected Components}
\label{sec:protected}

Every proposal passes two stages before it can alter the state used in production queries. The replay check toggles the proposal on and off on the current learning batch and requires that the net change in correct predictions be non-negative. The validation check evaluates the proposal as an A/B comparison on 100 images drawn from a pool sampled from the official held-out test split, disjoint from every reported test set and used for no other purpose (Appendix~\ref{app:valexam}), and rejects any proposal that lowers the correct count by more than one image.

The acceptance test is itself a component that can fail, so it carries a health check: the replay baseline arm must reproduce at least 70\% of the predictions known to be correct in the current round, failing which the test outcome for that round is disregarded and proposals fall back to sandbox-only admission combined with usage-statistics retirement. This check responds to a real incident (\S\ref{sec:failures}) in which a five-character argument-order bug in the scorer made both arms tie and every proposal pass, silently, for a day. Three rollback mechanisms surround the test, each isolated experimentally: an accuracy-drop rollback that restores the best snapshot when the test score falls by more than 3 points; a best-state restore that tracks the best validation-scored state and reinstates it at the end; and a batch interaction check that rejects an accepted batch when either class recall changes markedly, responding to the operating-point reversal failure mode of \S\ref{sec:failures}.

The set of protected components is fixed throughout the study. The model weights, the official system prompt, the scoring and parsing code, the learning-stream schedule, and all evaluation sets lie outside the loop's write access. The directive channel may be frozen, as in our default configuration, rewritten in its entirety, or patched as an enumerable rule list in which each rule is at most 40 words, individually tested and retractable, with at most six rules active. This three-way comparison follows the bounded, auditable-evolution design of \citet{shi2026sharp}, which constrains the evolvable surface of an agent in a high-stakes domain to individually testable, reversible artifacts.

\section{Experimental Setup}
\label{sec:setup}

\paragraph{Base detector and deployment simulation.}
The frozen base detector is MedForge-Reasoner \cite{chen-etal-2026-medforge}, namely Qwen3-VL-8B-Instruct \cite{qwen3vl} after SFT on 50{,}000 forgery-aware reasoning chains and group-sequence policy optimization on 10{,}000 images. MedForge-90K is officially split 5:1:3 into 50{,}000 SFT, 10{,}000 RL, and 30{,}000 held-out test images at a 1:1:1 real:implant:remove balance; all learning batches in this study draw from the 60{,}000-image training portion, and the validation pool and the final evaluation draw exclusively from the 30{,}000-image held-out portion. The detector is served identically across all runs by vLLM replicas with greedy decoding and the official system prompt. Self-improvement runs for 20 rounds. Each round presents a fresh learning batch of 40 newly drawn labeled images, balanced between fake and real and distributed over brain MRI, chest radiographs, and fundus photographs in fixed proportion; from round 2 onward, half of each batch receives one of five light distortions (two Gaussian-noise levels, two JPEG quality levels, and mild blur) to approximate the condition of images as they arrive in practice. Learning batches never repeat and never overlap any test set.

\paragraph{Test sets.}
All evaluation uses four 200-image sets scored in the final state, with the main set additionally scored in every round; the main, resample, and unseen-generator sets are mutually disjoint, and the distorted set is a transmission-degraded derivative of the main set.\footnote{The resample set was later found to draw 145 of its 200 images from the training pool; this inflates absolute scores on that set but leaves relative comparisons and all other sets unaffected, and we treat the final evaluation as the definitive measurement (Appendix~\ref{app:data}).} The \textbf{main} set contains 100 real images distributed 34/33/33 over the three modalities and 100 fake images from 10 generators with a 42/58 implant-to-remove ratio. The \textbf{resample} set is a same-distribution resample. The \textbf{unseen-generator} set contains 100 fake images from a generator$\times$modality combination absent from the main set and nearly absent from the learning stream, together with 100 real images. The \textbf{distorted} set applies three transmission distortions to the main set: large-factor resizing, print-and-rescan, and screen capture. The \textbf{final evaluation}, held out from all training images, scales the protocol to 2,000 clean and 2,000 distorted images (1,000 real in modality proportion; 10 generators $\times$ 100 images), drawn only from the official held-out test split and verified programmatically to have zero overlap with all training images and with the sets above. Appendix~\ref{app:data} lists the generators and distortion parameters, and Appendix~\ref{app:valexam} documents the validation pool used by the acceptance test.

\paragraph{Metrics.}
We report accuracy on each set together with real-image recall, the fraction of authentic images correctly passed, and fake-image recall, the fraction of forgeries correctly identified. The four-set average serves as the primary deployment metric, and per-set values are reported throughout. A 200-image set carries a 95\% confidence interval of roughly $\pm$5 points, whereas the final evaluation narrows this to $\pm$1.6; the principal comparisons therefore use the final evaluation.

\paragraph{Registered configurations.}
We study 49 single-switch configurations in four pre-registered batches, listed in full in Appendix~\ref{app:registry}: the core stores and their protections; eleven single mechanisms on a fixed base; a re-test of the mechanisms with positive effects on the strongest unprotected base, with new memory and admission mechanisms and seed replications; and the batch interaction check together with the recommended combination. Every configuration differs from its named base by exactly the listed switch.

\paragraph{Infrastructure and cost.}
The detector is served by four to eight identical vLLM replicas with per-variant concurrency 8, and no training compute is used anywhere in the study, whose total cost is below that of a single SFT run of the base detector. The external-reflector configurations use a strong instruction-tuned LLM (GLM-5.3-Flash) in that role only.

\section{Results}
\label{sec:results}

\begin{table*}[t]
\centering
\resizebox{0.92\textwidth}{!}{%
\setlength{\tabcolsep}{3.8pt}\renewcommand{\arraystretch}{0.98}
\begin{tabular}{l|ccccc|c|cc}
\toprule
\multirow{2}{*}{\textit{Configuration}} & \multicolumn{5}{c|}{Four test sets (200 images each, final state)} & \multirow{2}{*}{$\Delta$Avg} & \multicolumn{2}{c}{Final evaluation (2,000 each)} \\
\cmidrule(lr){2-6}\cmidrule(lr){8-9}
 & Main & Resample & Unseen-gen & Distorted & Avg. & & Clean & Distorted \\
\midrule
\rowcolor{blue!10} \multicolumn{9}{l}{\textit{Training-time progression (MedForge; weights then frozen)}} \\
Qwen3-VL-8B-Instruct (zero-shot) & \todo{} & \todo{} & \todo{} & \todo{} & \todo{} & --- & --- & --- \\
\;\;+ SFT (forgery-aware reasoning chains) & \todo{} & \todo{} & \todo{} & \todo{} & \todo{} & --- & --- & --- \\
\;\;+ SFT + RL (MedForge-Reasoner; no learning) & 79.5 & 79.0 & 77.0 & 64.5 & 75.0 & --- & 76.5 & 59.0 \\
\midrule
\rowcolor{blue!10} \multicolumn{9}{l}{\textit{Deployment-time self-improvement (this work; 20 rounds, weights frozen)}} \\
\;\;+ Experience (external reflector) & 84.5 & 85.5 & 85.0 & 72.0 & 81.8 & +6.8 & 83.8 & 65.8 \\
\;\;+ Self-reflection \& toolbox & 80.0 & 79.5 & 80.5 & 70.0 & 77.5 & +2.5 & --- & --- \\
\;\;+ Reading calibration & 86.5 & 84.5 & \underline{90.0} & 68.0 & 82.3 & +7.3 & 87.4 & 65.3 \\
\;\;+ Directive freeze$^{\dagger}$ & \textbf{87.0} & 86.5 & 87.5 & 73.0 & 83.5 & +8.5 & 85.4 & 66.1 \\
\;\;+ Ensemble reflection$^{\ddagger}$ & 86.5 & 87.5 & 88.0 & 75.5 & \textbf{84.4} & \textbf{+9.4} & \textbf{87.9} & \textbf{70.5} \\
\;\;+ Two-sided reflection$^{\ddagger}$ & 85.0 & \textbf{88.0} & 89.5 & 75.0 & \textbf{84.4} & \textbf{+9.4} & 86.9 & 65.7 \\
\;\;+ Case-based memory$^{\dagger}$ & 86.0 & 84.0 & 88.0 & \textbf{77.0} & 83.8 & +8.8 & --- & --- \\
\bottomrule
\end{tabular}}
\caption{\textbf{Main results.} Accuracy (\%) in the final state after 20 rounds. All self-improving rows share the same frozen detector, learning stream, and round budget. $^{\dagger}$built on the calibration variant; $^{\ddagger}$built on the calibration+freeze+rollback variant. Bold marks the best value per column, and underline marks the second best. The final evaluation is complete for six configurations; the self-reflection and case-based memory rows and the first two training-time rows are single-run or pending measurements and their final-evaluation cells are left blank accordingly.}
\label{tab:main}
\end{table*}

\subsection{Training-Time versus Deployment-Time Improvement}
\label{sec:ladder}

Table~\ref{tab:main} contrasts two sources of improvement: the rows above the divider rely on gradient updates before deployment, and the rows below rely on self-directed changes to the surrounding software afterwards. The training progression is highly effective on MedForge's own benchmark, reaching 99.2\% accuracy \cite{chen-etal-2026-medforge}, and this saturation explains why the remaining errors lie outside the benchmark: under our deployment-oriented protocol the same model averages 75.0\%, with real-image recall of 60\% and unseen-generator accuracy of 77.0\%, and every self-improving row operates on this identical frozen checkpoint.

The best configurations reach 84.4\%, a gain of 9.4 points obtained without any parameter update. On the 2,000-image final evaluation, ensemble reflection confirms the improvement at scale, raising clean accuracy from 76.5\% to 87.9\% and accuracy under transmission distortion from 59.0\% to 70.5\%, while real-image recall rises from 53.8\% to 89.3\% on clean images and from 22.1\% to 58.8\% on distorted images. Two-sided reflection shows the same pattern, at 86.9 and 65.7, with higher fake-image recall than ensemble reflection (94.0 and 88.4 vs.\ 86.4 and 82.2), the intended effect of the two-sided design on the balance between the two error directions.

\subsection{Where the Improvements Occur}
\label{sec:where}

Three observations refine the main result. First, the improvements address exactly the errors that remain after reinforcement learning: fake-image recall of the frozen detector is already 99\% on the main set, so every learning configuration instead raises real-image recall, from 60\% to 77--80\% for the calibration family, to 80\% for ensemble reflection, and to 91\% for case-based memory, reducing the false-alarm rate from 40\% to 20\%, and to 10\% with case-based memory. Second, the improvement generalizes: accuracy on the unseen-generator set rises from 77.0\% to between 88.0\% and 89.5\%, and ensemble and two-sided reflection both improve monotonically across the main, resample, and unseen-generator sets, indicating that the acquired capability is not memorization of the learning stream. Third, distorted images remain the most difficult condition but also show the largest gain: the distorted set is the lowest column for every configuration, yet the $+12.5$-point gain of case-based memory over the control is the largest single-set improvement in the table.

\subsection{Learning Curves Peak Early and Often Decline}
\label{sec:dynamics}

The 20-round trajectories (Figure~\ref{fig:curves}, Appendix~\ref{app:extra}) exhibit a common pattern. Every learning configuration reaches its maximum early, typically within the first ten rounds, after which the curves flatten, oscillate, or decline: averaged over the 18 learning configurations, the final state lies 3.0 points below the best round, five configurations lose 4.5 points or more, and four finish at or below the no-learning control despite strong intermediate performance, as in the case of second-opinion arbitration, which reaches 87.5\% at its best round and finishes at 78.5\%. The dynamics differ sharply with the protection in place. Single-perspective self-reflection rises to 84.0\% and then declines to 80.0\%; external experience without the calibration anchor oscillates; and the calibrated configurations sustain their plateau to the final round. The largest failure is invisible in overall accuracy, since one configuration finishes at an apparently acceptable 83.0\% while its real-image recall reaches 94\% and its fake-image recall only 72\%, a global reversal of the operating point that per-item checks could not detect (\S\ref{sec:failures}).

The acceptance test produces the largest single effect of any component. When a scoring bug silently disabled both of its stages for one batch of runs (\S\ref{sec:failures}), the same configuration that later reached 87.0\% finished at 69.5\%, a difference of 17.5 points in final accuracy attributable to the acceptance test alone.

\subsection{Effective and Ineffective Mechanisms}
\label{sec:works}

Table~\ref{tab:mechanisms} (Appendix~\ref{app:fourset}) isolates mechanisms under the single-switch protocol. Panel A traces the protection chain: converting raw tool readings into calibrated evidence contributes $+6.5$ points, freezing directive rewrites in their entirety adds a nominal $+0.5$ on the mean while eliminating the oscillation mode entirely, and adding accuracy-drop rollback in addition reduces final accuracy by 4.5 points once admission is already trustworthy, as it discards learning accumulated after the snapshot and caps performance at an early peak. Panel B shows the same cost of protection among single-switch additions, where several intuitive mechanisms underperform: tournament selection ($-6.0$), balanced per-class admission ($-5.0$), counterfactual reflection ($-5.0$), and second-opinion arbitration ($-4.0$) all impose stricter admission criteria or introduce additional evaluators into the decision, and all reduce final accuracy. The admission budget performs worst at $-24.0$, as a harmful round-one entry reversed the system's operating point and the budget then restricted the corrections that would have repaired it: a conservative mechanism applied at the wrong point blocks recovery rather than preventing damage.

The positive mechanisms act on three distinct stages of the loop, namely interpreting readings (calibration), supplying proposals (ensemble, two-sided, and red-team reflection), and filtering admission (dual validation), so the best final configurations in Table~\ref{tab:main} combine one mechanism from each group. A third panel re-tests eight mechanisms on the freeze base (Appendix~\ref{app:fourset}); its differences are attenuated, as the single run of the freeze base does not replicate across seeds (\S\ref{sec:seeds}), and the informative signal is again the ordering within the panel, in which case-based memory ranks highest. The composition reported in Table~\ref{tab:main} was selected to cover distinct stages on a protected base.

\section{Analysis}
\label{sec:analysis}

Having established that the loop improves the detector, we now examine which components are responsible and why the procedure sometimes fails, measuring the causal contribution of the self-written tools, testing whether the gains replicate across seeds, and characterizing the failure modes that the 49 trajectories expose.

\subsection{Effect of Self-Written Tools}
\label{sec:tools}

As tools are injected into the detector's context, their value is confounded with the experience entries that accompany them. Disabling the self-written tools in each frozen final state removes this confound (Appendix~\ref{app:tools}, Table~\ref{tab:selftool_full}): ten of the eleven measured configurations lose accuracy without their own tools, by 4.3 points on average and up to 10.0, and in most cases the loss is concentrated on the real-image side, with real-image recall falling by up to 15 points. The tools therefore supply calibrated evidence of authenticity that the detector otherwise lacks, although two configurations show the opposite pattern, in which their tools instead account for fake-image recall. Across the study the detector submitted 139 tool proposals, of which 104 were admitted, spanning 82 distinct tools repaired through 307 sandbox cycles, with code lengths between 21 and 85 lines.

The lifecycle of a tool is more informative than the number of tools. Figure~\ref{fig:lifecycle} follows one tool from proposal through sandbox failure, self-repair, deployment, and self-retirement. Without the discrimination table, self-written tools tend toward failure modes that our content audit quantifies: 61\% of implementations emit output from which no reading can be parsed, 26\% embed hardcoded numbers while computing unrelated quantities, and 22\% assume that the target lies at the image center (Appendix~\ref{app:audit}). One 60-line verifier applied its fabricated ``rib angles too dense, therefore fake'' rule to every image, yet passed every check, as its printed text was itself a persuasive instruction to report a forgery. A tool must therefore be evaluated on its numeric measurements rather than on the text it prints.

\subsection{Seed Replication Distinguishes Mechanisms from Random Variation}
\label{sec:seeds}

We re-ran two configurations with fresh learning streams. The calibration+freeze configuration, which produced the single best result in the study at 87.0, averages 79.3 across three seeds (87.0, 76.5, and 74.5; Table~\ref{tab:fourset}), a value statistically indistinguishable from the no-learning control, so its apparent advantage reflected the particular errors contained in this specific learning stream. Ensemble reflection averages 84.0 (86.5 and 81.5), a genuine improvement of 4.5 points over the control. Every principal result in this paper is accordingly reported on ensemble reflection, the only configuration replicated across seeds, or explicitly marked as single-seed (two-sided reflection and case-based memory). We regard seed replication as a methodological requirement for self-improvement claims, since a 20-round learning trajectory contains at least as many sources of variance as a training run, and several mechanisms in Table~\ref{tab:mechanisms} would not have been supported by a single-seed experiment.

\subsection{A Failure Taxonomy for Deployment-Time Self-Improvement}
\label{sec:failures}

Each failure mode below was observed and measured among the 49 trajectories; Figure~\ref{fig:failures} (Appendix~\ref{app:extra}) summarizes them.

\paragraph{I. Early gains that decline.}
The typical trajectory accumulates gains for 4 to 10 rounds and then declines as the supply of learnable error types is exhausted and new proposals begin to repeat or add noise (\S\ref{sec:dynamics}); one early run re-proposed the same experience entry (the absence of the corpus callosum) across eight consecutive directive versions. Diversity in the proposal supply, provided by ensemble reflection, two-sided reflection, and case-based memory, together with the retirement of stale entries, are the effective countermeasures.

\paragraph{II. Operating-point reversal from a batch of changes.}
After round 10 admitted seven changes in a single batch, each of them individually safe, the external-reflector-with-toolbox configuration reversed its operating point to 94\% real-image and 72\% fake-image recall, and the ten subsequent rounds did not recover. Once reversed, the system also scored near zero on the validation split in both arms of every later test, so the acceptance test loses discriminating power exactly when the system is already malfunctioning. Two safeguards were built and verified in response. The verifier health check requires the off-arm to reproduce at least 70\% of the predictions known to be correct in the current round, failing which the round's test is distrusted, and the batch interaction check rejects the entire accepted batch when either class recall changes markedly, producing zero false rejections across 40 rounds of checks. A complete description of this incident appears in Appendix~\ref{app:failnar}.

\paragraph{III. Single-run variability.}
As \S\ref{sec:seeds} shows, the best result of a single trajectory does not replicate across seeds. Since single-run gains remain the norm in this literature, this failure mode may be widespread.

\paragraph{IV. Silent verifier failure.}
A five-character argument-order bug in the acceptance test's scorer caused both arms to tie on every proposal, so every proposal passed for a full batch of runs, while the main evaluation pipeline, which uses different code, continued to report unaffected scores. These runs constitute an unplanned ablation of the acceptance test, costing the identical configuration 17.5 points of final accuracy (\S\ref{sec:dynamics}). The general principle is that the verifier is part of the attack surface of a self-improving system and must be externally checked, redundant, and simple enough to audit (Appendix~\ref{app:failnar}).

The study supports six design guidelines for deployment-time self-improvement, each traceable to a measured comparison above; we state them with their evidence in Appendix~\ref{app:guidelines}.

\section{Discussion and Conclusion}
\label{sec:discussion}

\paragraph{Established findings.}
Within the taxonomy of \citet{chen2026rsisurvey}, \rsi{} supplies the safety-critical perception instance of the deployment-time $\times$ human-on-the-loop cell (\S\ref{sec:related}): a published SFT+RL base, 49 registered configurations, seed replication, and a 4,000-image final evaluation. Three findings generalize beyond deepfake detection. External signal is inexpensive: each admitted change costs about 140 labeled validation images, and the entire study consumes a fraction of the labels of a single SFT epoch. The verification hierarchy of the survey predicts where the gains appear, since our strongest mechanisms all convert cheap and noisy signals into evidence anchored on labeled data. Finally, the verifier itself is part of the attack surface: our two worst incidents were failures of the verification components rather than of learning. For deployed detection the broader point is safety: medical deepfake generation methods continue to evolve after any detector is deployed, and these results show that verified self-improvement maintains detection performance under such evolution, avoiding both the obsolescence of a frozen system and the risk of unverified adaptation.

\paragraph{Relation to retraining.}
Deployment-time self-improvement improved real-image recall, transfer to unseen generators, and robustness to transmission distortion; fake-image recall, at 99\% after reinforcement learning, decreased to the high 80s and low 90s as the operating point was rebalanced, remaining above 86\% in every principal configuration on the final evaluation. These are the components of the error structure that reflect missing evidence and miscalibrated thresholds rather than missing capability. We therefore expect the division of labor to persist, with gradient updates establishing the base perceptual ability and recursive self-improvement adapting and extending it as the deployment distribution shifts. Periodic lightweight retraining on the accumulated experience and tools lies outside the frozen-weight design of this paper and is the most evident direction for future work.

\paragraph{Limitations and ethics.}
All results use one 8B detector and one benchmark family, so magnitudes will differ elsewhere. The learning stream draws labeled images from the training pool, and a deployment would require some verified label source, such as adjudicated clinical flags. The contamination of the resample set (\S\ref{sec:setup}) means that absolute four-set averages rest on the other three sets and on the final evaluation, which is complete for the control and the five leading self-improving configurations; the recommended four-mechanism combination was still running at submission time, and two-sided reflection and case-based memory lack seed replications. We claim only bounded self-refinement: the loop improves its surrounding software against a fixed evaluator and does not modify its own improvement mechanisms. All improvements concern the defensive side, and nothing in the loop generates or improves forgeries; the governance questions raised by a self-improving clinical decision system are partially addressed through protected verifiers, auditable trajectories, and individual acceptance of each change.

\paragraph{Conclusion.}
A frozen medical deepfake detector can improve itself substantially at deployment time, from 75.0\% to 84.4\% across four test sets and by 11 points on the 4,000-image final evaluation, under four conditions: every persistent change passes an independent-validation acceptance test, tool readings are calibrated rather than accepted as reported, write access is restricted to enumerable artifacts, and claims are confirmed by seed replication. The failure modes that remain, namely operating-point reversals, silent verifier failures, and single-run variability, are now characterized and paired with safeguards, and we release all artifacts needed to re-measure them.

\bibliography{refs}

\appendix
% Appendix figures are declared up front so LaTeX can place them into the
% text pages from the start; declaring them at their discussion points made
% the whole float queue drain onto the last pages, leaving half-empty columns.
\begin{figure}[t]
\centering
\includegraphics[width=0.93\linewidth]{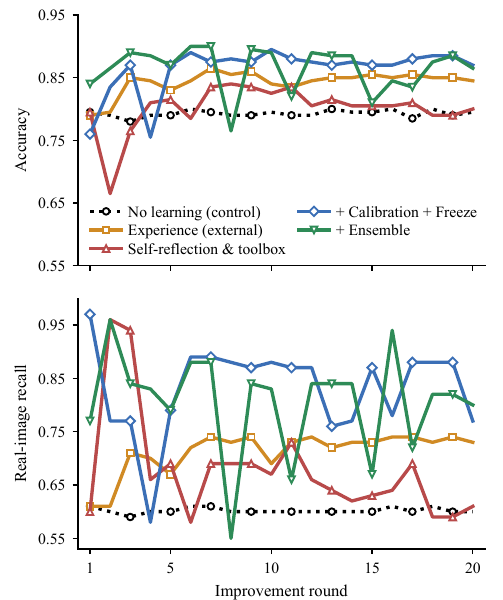}
\caption{\textbf{Learning dynamics over 20 rounds on the main test set.} The upper panel shows accuracy and the lower panel real-image recall. The frozen control remains constant at 79.5\% and 60\%. Configurations without calibration improve early in the run and subsequently decline, whereas the calibrated configurations maintain their improvement through the final round. Fake-image recall, which varies far more strongly for the uncalibrated configurations, appears in Figure~\ref{fig:allcurves}.}
\label{fig:curves}
\end{figure}

\begin{figure*}[t]
\centering
\includegraphics[width=0.92\textwidth]{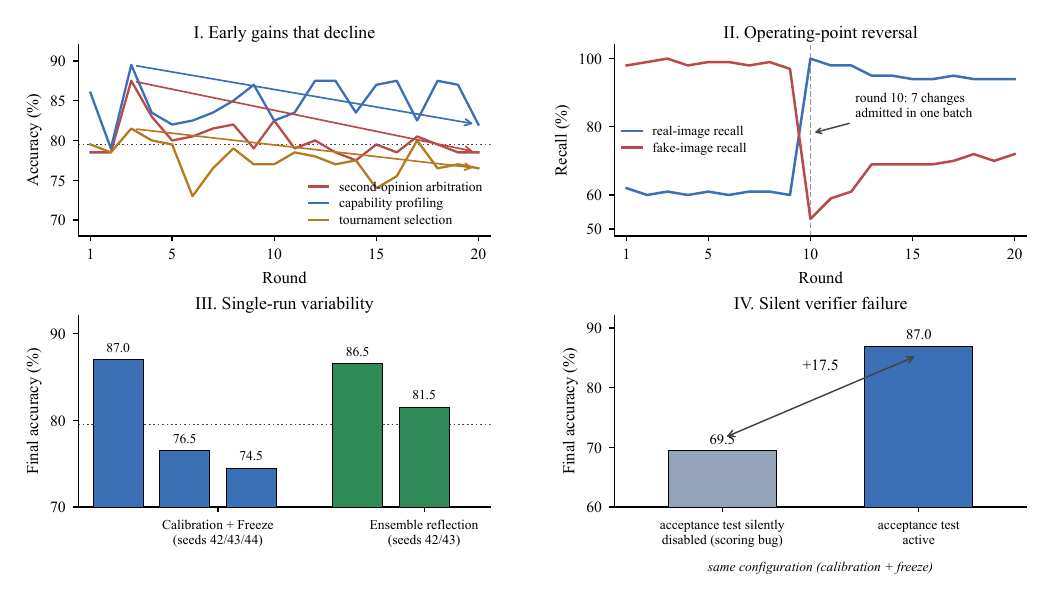}
\caption{\textbf{Four failure modes, plotted from measured trajectories.} Dotted lines mark the frozen control's final main-set accuracy (79.5\%) in panels I and III. (I) Early peaks that decline; the arrows mark the interval from peak to final value. (II) A batch of seven admitted changes reverses the operating point in round 10, and no subsequent round reverts it. (III) The best single run of the calibration+freeze configuration does not replicate, whereas ensemble reflection does. (IV) With the acceptance test disabled by a defect in the scoring code, the same configuration finishes 17.5 points lower.}
\label{fig:failures}
\end{figure*}

\begin{figure*}[t]
\centering
\includegraphics[width=0.92\textwidth]{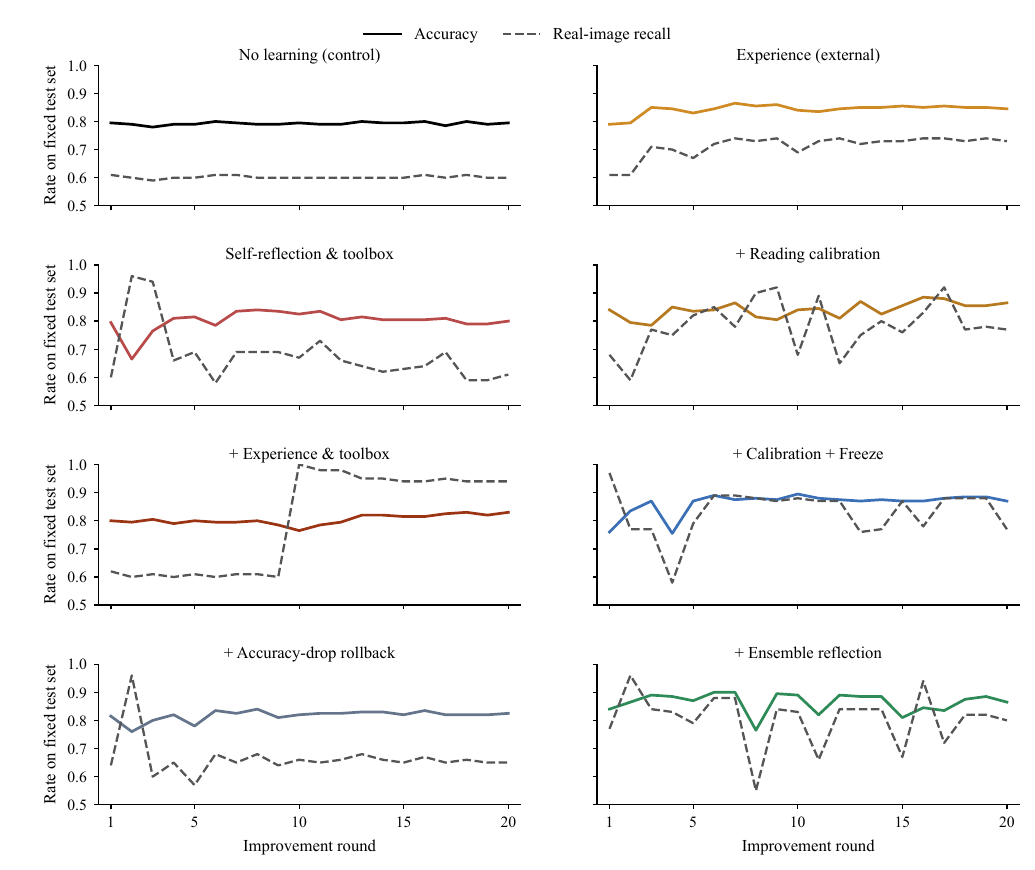}
\caption{\textbf{All 20-round trajectories} (accuracy and real-image recall) for eight representative configurations.}
\label{fig:allcurves}
\end{figure*}

\begin{figure}[t]
\centering
\includegraphics[width=0.93\linewidth]{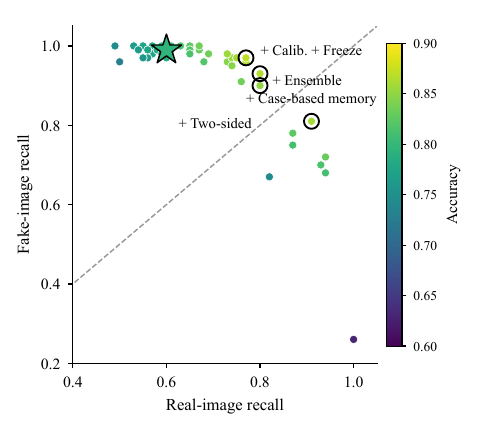}
\caption{\textbf{Error structure of all 47 completed configurations} (final state, main test set): real-image recall versus fake-image recall, colored by accuracy. The frozen control (star) lies at the saturated-fake / weak-real corner; every learning configuration shifts toward higher real recall. The operating-point-reversal configurations (bottom right) increase one recall at the expense of the other.}
\label{fig:asymmetry}
\end{figure}

\section{Registered Configuration Registry}
\label{app:registry}

Table~\ref{tab:registry} lists all 49 pre-registered configurations with their final main-set accuracy and class recalls. Mechanism descriptions are cumulative relative to the stated base; every pair (base, variant) differs by exactly the listed single modification. Batches: B1 = v1--v8 (core channels and protections), B2 = v9--v19 (one mechanism each on the fully protected base v8), B3 = v20--v45 (recombination on the rollback-free base, new mechanisms, seed replication), B4 = v46--v49 (batch-level admission check, recommended combination).

\begin{table*}[h]
\centering
\small
\setlength{\tabcolsep}{4pt}\renewcommand{\arraystretch}{1.05}
\begin{tabular}{l|l|ccc|l}
\toprule
\textit{ID} & \textit{Mechanism (cumulative)} & \textit{Main} & \textit{RealR} & \textit{FakeR} & \textit{Batch} \\
\midrule
v1 & No learning (frozen control) & 79.5 & 60 & 99  & B1 \\
v2 & Experience via external reflector (classic) & 84.5 & 73 & 96  & B1 \\
v3 & Self-reflection + autonomous toolbox & 80.0 & 61 & 99  & B1 \\
v4 & v3 + tool-reading calibration & 86.5 & 77 & 96  & B1 \\
v5 & External reflector + autonomous toolbox & 83.0 & 94 & 72  & B1 \\
v6 & v4 + directive freeze & 87.0 & 77 & 97  & B1 \\
v7 & v3 + accuracy-drop rollback (rollback at $>$3-pt drop) & 79.0 & 60 & 98  & B1 \\
v8 & v4 + freeze + accuracy-drop rollback (full stack) & 82.5 & 65 & 100  & B1 \\
v9 & v8 + second-opinion arbitration at low confidence & 78.5 & 57 & 100  & B2 \\
v10 & v8 + counterfactual reflection framing & 77.5 & 55 & 100  & B2 \\
v11 & v8 + dual validation splits & 85.5 & 74 & 97  & B2 \\
v12 & v8 + ensemble reflection (3 perspectives) & 86.5 & 80 & 93  & B2 \\
v13 & v8 + balanced per-class admission & 77.5 & 57 & 98  & B2 \\
v14 & v8 + capability profiling + targeted sampling & 82.0 & 65 & 99  & B2 \\
v15 & v8 + red-team audit of memory & 86.0 & 75 & 97  & B2 \\
v16 & v8 + tournament selection of proposals & 76.5 & 56 & 97  & B2 \\
v17 & v8 + self-consistency voting & 81.0 & 87 & 75  & B2 \\
v18 & v8 + calibrated confidence v2 (Wilson) & 83.0 & 67 & 99  & B2 \\
v19 & v8 + two-sided reflection (both error sides) & 85.0 & 80 & 90  & B2 \\
v20 & v6 + ensemble reflection & 79.5 & 59 & 100  & B3 \\
v21 & v6 + red-team audit & 81.5 & 63 & 100  & B3 \\
v22 & v6 + dual validation splits & 83.5 & 67 & 100  & B3 \\
v23 & v6 + two-sided reflection & 83.5 & 76 & 91  & B3 \\
v24 & v6 + ensemble + red-team & 82.5 & 87 & 78  & B3 \\
v25 & v4 + ensemble reflection & 78.5 & 60 & 97  & B3 \\
v26 & v6 + case-based memory (30 verified errors) & 86.0 & 91 & 81  & B3 \\
v27 & v6 + entry distillation every 5 rounds & 74.5 & 49 & 100  & B3 \\
v28 & v6 + forgetting (unused entries retire) & 82.0 & 68 & 96  & B3 \\
v29 & v6 + purging (counterfactual re-check) & 80.5 & 61 & 100  & B3 \\
v30 & v6 + toolbox subset search & 81.5 & 63 & 100  & B3 \\
v31 & v6 + plateau curriculum (difficulty step-up) & 84.5 & 74 & 95  & B3 \\
v32 & v6 + tool discrimination pre-check & 81.5 & 65 & 98  & B3 \\
v33 & v5 + calibration + freeze & 79.5 & 60 & 99  & B3 \\
v34 & v6 + admission budget (12 active, 2/round) & 63.0 & 100 & 26  & B3 \\
v35 & v6 + best-state restore (validation-selected) & 83.5 & 69 & 98  & B3 \\
v36 & v6 + entry-level admission threshold & 81.0 & 94 & 68  & B3 \\
v37 & v6 + 4-seat ensemble incl. audit seat & 73.0 & 50 & 96  & B3 \\
v38 & v6 + tournament with bench (losers suspended) & 78.5 & 58 & 99  & B3 \\
v39 & v4 + directive patches (rule list, $\leq$6) & 80.0 & 60 & 100  & B3 \\
v40 & v6 + field-statistics labels on entries & 76.5 & 53 & 100  & B3 \\
v41 & v6 + calibration sliding window (10 rounds) & 78.5 & 59 & 98  & B3 \\
v42 & v6 + admission budget + ensemble & 77.5 & 56 & 99  & B3 \\
v43 & v6, learning-stream seed 43 & 76.5 & 54 & 99  & B3 \\
v44 & v6, learning-stream seed 44 & 74.5 & 82 & 67  & B3 \\
v45 & v12, learning-stream seed 43 & 81.5 & 93 & 70  & B3 \\
v46 & v3 + batch interaction check & 76.0 & 55 & 97  & B4 \\
v47 & v6 + batch interaction check & 85.5 & 73 & 98  & B4 \\
v48 & v6 + ensemble + two-sided + case-based memory (recommended) & run. & run. & run.  & B4 \\
v49 & v48 + accuracy-drop rollback & run. & run. & run.  & B4 \\
\bottomrule
\end{tabular}
\caption{\textbf{Complete registry of the 49 registered configurations.} Main-set final accuracy (\%) and real/fake-image recall after 20 rounds. ``run.'' = still completing at submission. Configurations discussed in the main text: control v1; experience channel v2; self+toolbox v3; calibration v4; freeze v6; rollback stack v8; ensemble v12; two-sided v19; case-based memory v26; seeds v43--v45; recommended combination v48/v49.}
\label{tab:registry}
\end{table*}

\begin{figure}[t]
\centering
\includegraphics[width=0.93\linewidth]{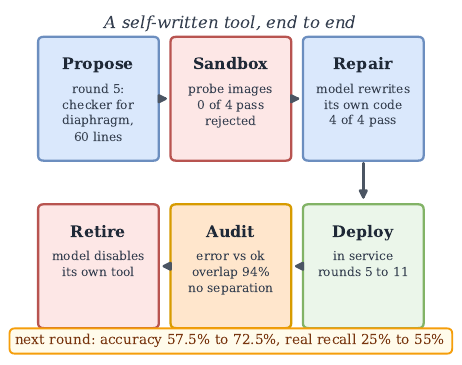}
\caption{\textbf{Lifecycle of a self-written tool} (a diaphragm-consistency verifier). The proposal fails all four sandbox probes; the reflector repairs its own code and the tool enters service. Six rounds later the discrimination table shows no separation, and the model disables its own tool; accuracy recovers from 57.5\% to 72.5\% in the following round (real-image recall 25\%$\rightarrow$55\%). The write-repair-deploy-audit-retire cycle runs without human intervention.}
\label{fig:lifecycle}
\end{figure}
\needspace{8\baselineskip}
\section{Harness Implementation Details}
\label{app:implementation}

\paragraph{Detection query.}
The official MedForge-Reasoner system prompt is used verbatim; all evolvable content (directive, retrieved experience entries, tool readings, case precedents) is rendered into the user turn. The verdict is parsed from a fixed conclusion line by scoring code outside the loop's write access; ambiguity resolves to a documented default.

\paragraph{Experience entry schema.}
Each entry consists of: \textit{trigger} (modality $\times$ generator-family $\times$ distortion conditions, with explicit fallback to general entries when none match), \textit{cue} (the visual finding to seek, phrased as evidence rather than verdict), \textit{caution} (the error it protects against), field statistics (\textit{times-correct}/\textit{times-incorrect} counts over injections), and a status field (probation / active / retired). Retirement triggers: sustained negative field statistics, red-team nomination with net-positive protection, $N$-round non-use (forgetting), or counterfactual re-check (purging).

\paragraph{Tool contract and sandbox.}
Tools are single self-contained files, imports limited to Pillow and NumPy; execution budget 15\,s; output $\leq$600 characters and must contain at least one numeric reading; exceptions forbidden (return an error string instead). The sandbox executes each candidate on four probe images (two real, two fake, one distorted); failures are returned verbatim to the reflector for repair. Each tool proposal specifies an action (\texttt{new}/\texttt{modify}/\texttt{enable}/\texttt{disable}) and, for \texttt{new}, the complete source file.

\paragraph{Reading calibration.}
For each tool and each numeric output field, the harness maintains reading histograms over historical images with known labels (capped window). At query time a raw reading $r$ is rendered as percentile pairs $(P_{\mathrm{fake}}(r), P_{\mathrm{real}}(r))$ with a direction label when the historical distributions separate (quantile gap threshold) and \textsc{no signal} otherwise. The Wilson lower bound variant (calibrated confidence v2) additionally down-weights small samples.

\paragraph{Acceptance test.}
Stage 1 (replay): A/B on the round's 40 learning images, net correct-change $\geq 0$. Stage 2 (validation): A/B on 100 images sampled from the 200-image validation pool, correct-count drop $\leq 1$. Verifier-health check: the off-arm must reproduce $\geq 70\%$ of this round's known-correct answers; on failure the acceptance-test outcome is disregarded for that round, and proposals instead require sandbox acceptance together with field-statistics-based retirement. Dual-validation splits require both disjoint 100-image halves to pass. Batch-level joint check (v46/v47): after per-item admission, the whole accepted batch is applied and either class recall on the validation split is compared with the round start; a drop exceeding 25 percentage points on either side causes the batch to be rejected and the round-start state to be restored.

\paragraph{Reflection prompts (summary).}
The self-reflection dialogue omits the detection prompt and has two stages: (1) the model inspects the misclassified images (and a sample of correct ones) and drafts a free-form diagnosis; (2) it emits strictly-parsed JSON proposals. The focus-rotation schedule cycles through six diagnostic themes and presents the reflector with its own directive history. Ensemble reflection runs three independent stage-2 sessions (different temperatures and perspective instructions) and merges non-conflicting proposals. Two-sided reflection presents false alarms and missed detections in parallel and requires each proposal to declare its target direction and its non-harm evidence for the opposite direction. The external reflector (experience-only configurations) receives the same error summary in text form. Verbatim prompt skeletons appear in Appendix~\ref{app:prompts}.

\paragraph{Infrastructure.}
Detector: the MedForge-Reasoner model (Qwen3-VL-8B-Instruct after SFT+GSPO), served by four to eight vLLM replicas (max length 32{,}768, greedy). External reflector, used only in the three experience-channel configurations: GLM-5.3-Flash (role boundaries in Appendix~\ref{app:llm}). Per-variant concurrency 8, six-fold timeout retry, multi-endpoint failover. A 20-round configuration completes in 2--7 hours of wall-clock time depending on the calibration variant; per-round fixed-test scoring takes 6.4--7.2 minutes (200 images; Appendix~\ref{app:cost}).

\section{Hyperparameters and Acceptance-Test Settings}
\label{app:hyperparams}

Table~\ref{tab:hyper} and Table~\ref{tab:hyperllm} collect the fixed settings shared by all registered configurations. Values are read from the released per-variant configuration files and harness code; single-switch configurations change exactly one of these values relative to their stated base.

\begin{table}[h]
\centering
\small
\setlength{\tabcolsep}{2.5pt}\renewcommand{\arraystretch}{1.0}
\begin{tabularx}{\linewidth}{@{}p{0.36\linewidth}p{0.14\linewidth}>{\raggedright\arraybackslash}X@{}}
\toprule
\textit{Parameter} & \textit{Value} & \textit{Notes} \\
\midrule
\multicolumn{3}{@{}l}{\textit{Loop schedule}} \\
Rounds per configuration & 20 & fresh learning batch each round \\
Learning images per round & 40 & 20 fake + 20 real, no repeats \\
Degraded share of learning images & 0.5 & from round 2; kinds rotate (Table~\ref{tab:perturb}) \\
\midrule
\multicolumn{3}{@{}l}{\textit{Context rendering}} \\
Experience entries per query & top-5 & trigger specificity first, fallback to general entries \\
Rendered prompt cap & 4{,}000 chars & directive + entries + tool readings \\
Tool-reading text cap & 700 chars & per tool, after calibration annotation \\
Image long side & 1{,}024 px & \\
Detector decoding & greedy & max 2{,}048 tokens \\
\midrule
\multicolumn{3}{@{}l}{\textit{Acceptance test}} \\
Stage-1 replay set & $\leq$40 images & all round errors + equal correct sample \\
Stage-1 pass rule & $\geq$0 & net correct-change on replay \\
Stage-2 validation size & 100 & fixed subset of the 200-image pool (Appendix~\ref{app:valexam}) \\
Stage-2 pass rule & $\leq$1 & correct-count drop on validation \\
Dual-validation split & odd/even & disjoint halves, both must pass (v11, v22) \\
Verifier health check & $\geq$70\% & replay of known-correct answers; one retry, then admission by sandbox only \\
Batch interaction check & $\leq$25 pts & class-recall swing on 60 validation images, else batch restored (v46, v47) \\
\bottomrule
\end{tabularx}
\caption{\textbf{Fixed hyperparameters shared by the study (loop schedule, context rendering, acceptance test).} Single-switch mechanisms modify one row (or introduce one mechanism not listed here); no configuration adjusts any of these values during its run.}
\label{tab:hyper}
\end{table}

\begin{table}[h]
\centering
\small
\setlength{\tabcolsep}{2.5pt}\renewcommand{\arraystretch}{1.0}
\begin{tabularx}{\linewidth}{@{}p{0.36\linewidth}p{0.14\linewidth}>{\raggedright\arraybackslash}X@{}}
\toprule
\textit{Parameter} & \textit{Value} & \textit{Notes} \\
\midrule
\multicolumn{3}{@{}l}{\textit{Reflection}} \\
Stage-1 diagnosis & $T{=}0.4$ & multimodal, free-form, 150--400 words \\
Stage-2 proposals & $T{=}0.2$ & strict JSON (Appendix~\ref{app:prompts}) \\
JSON repair retry & 1, $T{=}0.0$ & presents the parse error to the reflector \\
Tool self-repair & $\leq$3, $T{=}0.1$ & sandbox failures returned verbatim \\
Ensemble sessions & 3 & main $T{=}0.2$; missed-fakes seat $T{=}0.6$; false-alarm seat $T{=}0.85$ (4-seat variant adds measurement $T{=}0.7$ and audit $T{=}0.4$ seats) \\
Focus rotation & 6 themes & one per round, cyclically \\
Proposal quota per round & 3/1/3/2 & entries / directive patches / tool actions / retirements \\
\midrule
\multicolumn{3}{@{}l}{\textit{Memory}} \\
Live experience-entry cap & 40 & per-variant library \\
Directive rule list & $\leq$6 rules & $\leq$2 changes per round, each rule separately validated (v39) \\
Case-based memory & 30 & precedents, seeded with verified errors (v26) \\
\midrule
\multicolumn{3}{@{}l}{\textit{Serving}} \\
Per-variant concurrency & 8 & \\
API timeout / retries & 90\,s / 6 & multi-endpoint failover \\
\bottomrule
\end{tabularx}
\caption{\textbf{Fixed hyperparameters shared by the study (reflection, memory, serving).} Row meanings are identical to Table~\ref{tab:hyper}; no configuration tunes any of these values during its run.}
\label{tab:hyperllm}
\end{table}

\section{Reflection Prompt Structure}
\label{app:prompts}

In self-reflection configurations the reflector is the frozen detector itself, addressed in its normal question-answering mode without the forensic system prompt. Each round it receives the verified result summary, per-generator and per-distortion accuracy, error clusters, the live experience library with usage statistics, its own directive history, and a tool-readout table contrasting its error cases with correctly-judged cases from the same cluster, followed by the contrast images themselves. The skeletons below are abridged to the essential instructions; seat labels and JSON field names are renamed to match the terminology of this paper.

\paragraph{Stage 1: free-form diagnosis (temperature 0.4, with images).}
\begin{tcolorbox}[colback=gray!10, colframe=gray!50, title=Abridged stage-1 diagnosis prompt, breakable]
\footnotesize
\begin{verbatim}
TASK - write a free-form diagnosis (no
JSON, no special format, 150-400 words):
1. Look at the images: what visually
   separates the fakes from the reals
   here? Why did you misjudge the error
   cases? Be concrete about textures,
   anatomy, noise, artifacts.
2. Read the tool table: which readings
   carry signal (differ between WRONG
   and RIGHT cases)? Which are noise
   or misleading?
3. Decide what you need: a new measuring
   tool (what exactly should it
   quantify?), a fix to an existing
   tool (what change?), enabling or
   disabling, and/or a library lesson
   / directive tweak.
Write your honest reasoning; this will
not be shown to anyone but yourself.
\end{verbatim}
\end{tcolorbox}

\paragraph{Stage 2: strict JSON proposals (temperature 0.2).}
\begin{tcolorbox}[colback=gray!10, colframe=gray!50, title=Abridged stage-2 proposal schema, breakable]
\footnotesize
\begin{verbatim}
TASK - output this round's improvement
proposals as STRICT JSON only (first
char '{', last '}', no prose, no
markdown fences). Schema:
{"diagnosis_summary": "one line, MAX
 15 words",
 "templates": [{"trigger":
     {"generator":"*", "modality":"*",
      "type":"*", "perturb":"gauss_10"},
  "cue": "...", "caution": "..."}],
 "prompt_patch": {"rationale":"...",
      "new_directive":"..."} | null,
 "tool_actions": [{"action":"new_tool",
      "name":"snake_case",
      "description":"...",
      "code":"<COMPLETE file>"}],
 "retire_ids": ["EXP-0001"]}
EVIDENCE-ONLY OUTPUT: report measured
values with their context (e.g.
 "value=3.2 (natural range 0.8-2.1)").
Do NOT put a verdict in the tool
output - no "-> FAKE", no "-> REAL":
the reading is evidence for the
detector to weigh, and the calibration
layer adds the direction.
\end{verbatim}
\end{tcolorbox}
Malformed JSON triggers one automatic repair retry at temperature 0.0, in which the parse error is presented to the model; tool files rejected by the sandbox undergo up to three self-repair rounds at temperature 0.1 with the failure log attached.

\paragraph{Ensemble seats and the two-direction error list.}
\begin{tcolorbox}[colback=gray!10, colframe=gray!50, title=Abridged ensemble seat instruction and two-direction error list, breakable]
\footnotesize
\begin{verbatim}
[ensemble seat: missed-fakes specialist]
You are the missed-fakes specialist:
concentrate ONLY on FAKE images that
were wrongly cleared. What forensic
cue would have exposed them?

[two-direction error list, in the material]
Direction A - REAL images flagged as
FAKE (n this round): <cases>
Direction B - FAKE images cleared as
REAL (n this round): <cases>
RULE: 1. each lesson must name which
direction it aims to fix; 2. it must
state what evidence keeps it from
harming the opposite direction; 3. if
Direction A has >= 2 errors this
round, at least one proposal MUST
target A.
\end{verbatim}
\end{tcolorbox}
The false-alarm seat (temperature 0.85) mirrors the missed-fakes instruction on wrongly flagged real images; the optional measurement seat (0.7) is restricted to failure modes a deterministic analyzer could measure, and the audit seat (0.4) may only name suspect library entries for retirement. The system material additionally provides a list of fourteen measurement families (frequency-domain analysis, noise-residual statistics, error-level analysis, block-grid strength, symmetry, texture descriptors, cross-scale consistency, and others), offered as candidate directions rather than an exhaustive specification.

\section{Validation-Set Construction}
\label{app:valexam}

All admission decisions are scored on a validation pool that is disjoint from every reported test set, which addresses the main methodological concern of reusing one split for both learning and reporting. The pool is drawn from the official held-out test split of MedForge-90K by the same stratified procedure as the fixed main test set, with programmatic path-level exclusion of all 200 main-set images: fake images are drawn cyclically across the ten generators (ten images each), and real images fill the remaining half, giving 200 images (100 fake, 100 real; composition in Table~\ref{tab:valexam}). Each acceptance test evaluates a fixed, stride-sampled 100-image subset of the pool; the dual-validation configurations additionally split this subset into disjoint odd- and even-indexed halves, both of which must pass. The pool is used only for admission decisions (stage-2 validation, the batch interaction check, and toolbox subset search); the fixed test set is scored for reporting only and never enters any decision, and the four-set final evaluation uses separate resample, unseen-generator, and distorted splits.

\begin{table}[h]
\centering
\small
\setlength{\tabcolsep}{5pt}
\begin{tabular}{l|c|c}
\toprule
\textit{Validation-pool slice} & \textit{Fake} & \textit{Real} \\
\midrule
Total images & 100 & 100 \\
Fundus & 41 & 34 \\
Chest radiograph & 28 & 31 \\
Brain MRI & 31 & 35 \\
Edit-type fakes (edit / removal) & 46 / 54 & --- \\
Per generator & 10 each & --- \\
\bottomrule
\end{tabular}
\caption{\textbf{Composition of the 200-image validation pool}, computed from the released pool file. Fake images are balanced ten per generator; real images are drawn from the same three modalities as the test set. The pool is disjoint from the fixed test set and is used only for admission decisions.}
\label{tab:valexam}
\end{table}

\section{Data Composition and Degradation Protocol}
\label{app:data}

The fixed main test set contains 200 images (100 fake, 100 real). Fake images cover ten generators with ten images each (Table~\ref{tab:generators}); 58 are removal-type edits and 42 are other edits, split across modalities as fundus 39, chest radiograph 37, brain MRI 24. Real images are drawn 33/34/33 from chest radiograph, brain MRI, and fundus. The unseen-generator and resample evaluation sets follow the same stratification with disjoint generators and resampled real images, respectively. The resample set was later found to draw 145 of its 200 images from the training pool, that is, from images included in the detector's SFT training data; absolute scores on that set are inflated equally for every configuration. No other set is affected: the main, unseen-generator, and distorted sets and the final evaluation draw exclusively from images outside the detector's training data, and the final evaluation is therefore treated as the definitive measurement.

\begin{table}[h]
\centering
\small
\setlength{\tabcolsep}{5pt}
\begin{tabular}{l|l}
\toprule
\multicolumn{2}{c}{\textit{Generators of main-set fake images (10 images each)}} \\
\midrule
flux.1-dev & stable-diffusion-3.5-large \\
gemini & stable-diffusion-3.5-medium \\
gpt & stable-diffusion-inpainting \\
qwen & stable-diffusion-xl-1.0-inpainting-0.1 \\
seeddream & step1x-edit \\
\bottomrule
\end{tabular}
\caption{\textbf{The ten fake-image generators} of the main set, as named in the dataset metadata. Two are inpainting models, one is an instruction-edit model, and the remainder are full-image generators.}
\label{tab:generators}
\end{table}

Learning degradations approximate propagation loss on the learning stream only; transmission degradations define the distorted evaluation set. Both families are deterministic functions of the image and are cached, so every configuration receives byte-identical degraded inputs (Table~\ref{tab:perturb}).

\begin{table}[h]
\centering
\small
\setlength{\tabcolsep}{4pt}
\begin{tabularx}{\linewidth}{@{}l l X@{}}
\toprule
\textit{Kind} & \textit{Parameter} & \textit{Definition} \\
\midrule
\multicolumn{3}{@{}l}{\textit{Learning degradations}} \\
gauss\_10 & $\sigma{=}10/255$ & additive Gaussian noise \\
jpeg\_40 & quality 40 & JPEG re-compression \\
blur\_1 & radius 1.0\,px & Gaussian blur \\
gauss\_15 & $\sigma{=}15/255$ & additive Gaussian noise \\
jpeg\_25 & quality 25 & JPEG re-compression \\
\midrule
\multicolumn{3}{@{}l}{\textit{Transmission degradations}} \\
resize\_035 & factor 0.35 & bilinear rescale to 35\% of the original size, then back; saved JPEG 95 \\
printscan & q35 + 0.8\,px & JPEG 35, then Gaussian blur radius 0.8\,px; saved JPEG 90 \\
screen\_075 & 0.75 + q50 & rescale to 75\%, then JPEG 50 \\
\bottomrule
\end{tabularx}
\caption{\textbf{Degradation protocol.} Learning degradations are injected into the learning stream to make lessons robust to propagation loss; transmission degradations emulate low-resolution resharing, print-and-scan, and screen-capture or social-media recompression chains for the distorted evaluation set.}
\label{tab:perturb}
\end{table}

\section{Complete Per-Variant Four-Set Results}
\label{app:fourset}

\begin{figure}[h]
\centering
\includegraphics[width=0.93\linewidth]{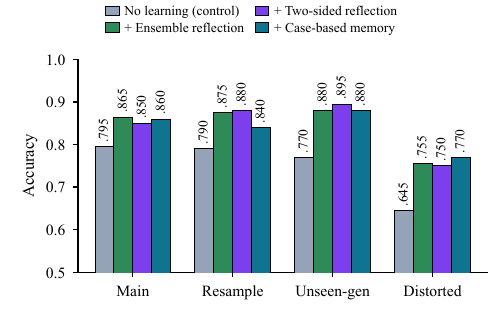}
\caption{\textbf{Per-set accuracy of the four leading configurations versus the frozen control.} Case-based memory is strongest under distortion (77.0); two-sided reflection is strongest on the resample and unseen-generator sets (88.0, 89.5), and ensemble and two-sided reflection both improve monotonically across main $\rightarrow$ resample $\rightarrow$ unseen-generator.}
\label{fig:foursets}
\end{figure}

Table~\ref{tab:fourset} reports the final-state four-set accuracy of every configuration with complete evaluations, and Figure~\ref{fig:foursets} plots the four leading configurations against the frozen control on each set.

\begin{table*}[h]
\centering
\small
\resizebox{\textwidth}{!}{%
\setlength{\tabcolsep}{5pt}\renewcommand{\arraystretch}{1.05}
\begin{tabular}{l|cccc|c|cc|cc}
\toprule
\textit{Config} & \textit{Main} & \textit{Resample} & \textit{Unseen} & \textit{Distorted} & \textit{Avg} & \textit{Peak (rd)} & \textit{Tail-5} & \textit{RealR} & \textit{FakeR} \\
\midrule
v1 control & 79.5 & 79.0 & 77.0 & 64.5 & 75.0 & 80.0 (6) & 79.4 & 60 & 99 \\
v2 experience-ext & 84.5 & 85.5 & 85.0 & 72.0 & 81.8 & 86.5 (7) & 85.0 & 73 & 96 \\
v3 self+tools & 80.0 & 79.5 & 80.5 & 70.0 & 77.5 & 84.0 (8) & 79.9 & 61 & 99 \\
v4 +calibration & 86.5 & 84.5 & 90.0 & 68.0 & 82.3 & 88.5 (16) & 86.8 & 77 & 96 \\
v5 ext+tools & 83.0 & 79.5 & 86.5 & 75.5 & 81.1 & 83.0 (18) & 82.4 & 94 & 72 \\
v6 +freeze & 87.0 & 86.5 & 87.5 & 73.0 & 83.5 & 89.5 (10) & 87.8 & 77 & 97 \\
v7 rollback only & 79.0 & 83.0 & 82.5 & 69.5 & 78.5 & 83.5 (5) & 81.4 & 60 & 98 \\
v8 full stack & 82.5 & 81.0 & 81.0 & 68.5 & 78.3 & 84.0 (8) & 82.4 & 65 & 100 \\
v9 +second-op & 78.5 & 77.0 & 78.5 & 64.0 & 74.5 & 87.5 (3) & 79.1 & 57 & 100 \\
v10 +counterfactual & 77.5 & 71.5 & 76.0 & 64.0 & 72.3 & 79.0 (7) & 76.8 & 55 & 100 \\
v11 +dual-val & 85.5 & 85.5 & 85.0 & 74.5 & 82.6 & 86.5 (5) & 85.0 & 74 & 97 \\
v12 +ensemble & 86.5 & 87.5 & 88.0 & 75.5 & \textbf{84.4} & 90.0 (6) & 86.1 & 80 & 93 \\
v13 +balanced & 77.5 & 76.5 & 78.5 & 67.0 & 74.9 & 80.5 (1) & 78.1 & 57 & 98 \\
v14 +profiling & 82.0 & 86.5 & 86.5 & 71.5 & 81.6 & 89.5 (3) & 85.3 & 65 & 99 \\
v15 +red-team & 86.0 & 86.5 & 85.0 & 69.0 & 81.6 & 86.0 (19) & 82.7 & 75 & 97 \\
v16 +tournament & 76.5 & 80.5 & 79.5 & 69.5 & 76.5 & 81.5 (3) & 77.1 & 56 & 97 \\
v17 +self-consistency & 81.0 & 83.5 & 80.0 & 70.5 & 78.8 & 85.5 (8) & 77.6 & 87 & 75 \\
v18 +calibrated confidence (variant 2) & 83.0 & 82.0 & 83.5 & 73.0 & 80.4 & 85.5 (10) & 82.6 & 67 & 99 \\
v19 +two-sided & 85.0 & 88.0 & 89.5 & 75.0 & \textbf{84.4} & 85.5 (19) & 83.2 & 80 & 90 \\
v26 (on v6) +case-mem & 86.0 & 84.0 & 88.0 & \textbf{77.0} & 83.8 & 87.5 (19) & 86.0 & 91 & 81 \\
\bottomrule
\end{tabular}}
\caption{\textbf{Four-set final-state results for all configurations with complete evaluations} (accuracy, \%). Peak = best round on the main set (round in parentheses); Tail-5 = mean of the last five rounds. The remaining 27 completed configurations are single-run and listed with main-set statistics in Appendix~\ref{app:registry}; their four-set evaluations follow the same protocol.}
\label{tab:fourset}
\end{table*}

\begin{table}[h]
\centering
\resizebox{\linewidth}{!}{%
\setlength{\tabcolsep}{2.6pt}\renewcommand{\arraystretch}{1.02}
\begin{tabular}{l|c|c}
\toprule
\textit{Configuration (cumulative chain)} & \textit{Main} & $\Delta$ \\
\midrule
\rowcolor{blue!8} \multicolumn{3}{l}{\textit{(A) Protection chain: what the acceptance era adds}} \\
Self-reflection \& toolbox & 80.0 & --- \\
\;+ reading calibration & 86.5 & $+6.5$ \\
\;\;+ directive freeze & 87.0 & $+0.5$ \\
\;\;\;+ accuracy-drop rollback & 82.5 & $-4.5$ \\
\midrule
\rowcolor{green!8} \multicolumn{3}{l}{\textit{(B) One mechanism on the rollback base (82.5)}} \\
\;+ ensemble reflection & \textbf{86.5} & $+4.0$ \\
\;+ red-team memory audit & 86.0 & $+3.5$ \\
\;+ dual validation splits & 85.5 & $+3.0$ \\
\;+ two-sided reflection & 85.0 & $+2.5$ \\
\;+ calibrated confidence (variant 2) & 83.0 & $+0.5$ \\
\;+ capability profiling & 82.0 & $-0.5$ \\
\;+ self-consistency voting & 81.0 & $-1.5$ \\
\;+ second-opinion arbitration & 78.5 & $-4.0$ \\
\;+ counterfactual reflection & 77.5 & $-5.0$ \\
\;+ balanced per-class admission & 77.5 & $-5.0$ \\
\;+ tournament selection & 76.5 & $-6.0$ \\
\bottomrule
\end{tabular}}
\caption{\textbf{Mechanism isolation} (main-set final accuracy, \%; one switch per row). Ensemble reflection gains $+4.0$ here yet $-7.5$ on the freeze base (Table~\ref{tab:panelc}), as the use of diverse reflectors triples the number of proposals and requires rollback protection.}
\label{tab:mechanisms}
\end{table}

\begin{table}[h]
\centering
\small
\setlength{\tabcolsep}{2pt}
\begin{tabular}{l|c|c}
\toprule
\textit{Eight mechanisms on the freeze base (87.0$^{*}$)} & \textit{Main} & $\Delta$ \\
\midrule
\;+ case-based memory & 86.0 & $-1.0^{*}$ \\
\;+ plateau curriculum & 84.5 & $-2.5^{*}$ \\
\;+ dual validation splits & 83.5 & $-3.5^{*}$ \\
\;+ two-sided reflection & 83.5 & $-3.5^{*}$ \\
\;+ best-state restore & 83.5 & $-3.5^{*}$ \\
\;+ ensemble reflection & 79.5 & $-7.5^{*}$ \\
\;+ entry distillation & 74.5 & $-12.5^{*}$ \\
\;+ admission budget & 63.0 & $-24.0^{*}$ \\
\bottomrule
\end{tabular}
\caption{\textbf{Panel C of Table~\ref{tab:mechanisms}}: eight mechanisms, one each on the freeze base. $^{*}$The base's 87.0 does not replicate across seeds (76.5/74.5; mean 79.3, \S\ref{sec:seeds}), so deltas measured against that single run are downward-biased.}
\label{tab:panelc}
\end{table}

\section{Seed Replication}
\label{app:seeds}

Replication changes only the learning stream: the fixed test set, the validation pool, and all harness settings are byte-identical across seeds, so differences measure the variance of the learning trajectory itself. Table~\ref{tab:seeds} lists the replications. The calibration+freeze configuration does not replicate (mean 79.3, statistically indistinguishable from the 79.5 control), whereas ensemble replication holds (mean 84.0, a 4.5-point improvement over the control); \S\ref{sec:seeds} discusses the consequences for \mbox{reporting}.

\begin{table}[h]
\centering
\small
\setlength{\tabcolsep}{2.5pt}
\begin{tabular}{l|c|ccc}
\toprule
\textit{Configuration} & \textit{Seed} & \textit{Main} & \textit{RealR} & \textit{FakeR} \\
\midrule
v6 calibration+freeze & 42 (base) & 87.0 & 77 & 97 \\
v43 = v6, new stream & 43 & 76.5 & 54 & 99 \\
v44 = v6, new stream & 44 & 74.5 & 82 & 67 \\
\quad mean over seeds & & 79.3 & & \\
\midrule
v12 ensemble reflection & 42 (base) & 86.5 & 80 & 93 \\
v45 = v12, new stream & 43 & 81.5 & 93 & 70 \\
\quad mean over seeds & & 84.0 & & \\
\bottomrule
\end{tabular}
\caption{\textbf{Seed replication} (main-set final accuracy, \%; final class recalls per run). The frozen control is 79.5. Only the learning stream is re-sampled; evaluation sets are unchanged.}
\label{tab:seeds}
\end{table}

\section{Complete Self-Tool Ablation}
\label{app:tools}

Table~\ref{tab:selftool_full} reports the enabled-versus-disabled comparison for every configuration whose self-written tools were measured; the main text discusses the pattern.

\begin{table}[h]
\centering
\small
\resizebox{\linewidth}{!}{%
\setlength{\tabcolsep}{2.8pt}\renewcommand{\arraystretch}{1.05}
\begin{tabular}{l|c|cc|cc|ccc}
\toprule
\textit{Config} & \textit{\#tools} & \multicolumn{2}{c|}{\textit{With}} & \multicolumn{2}{c|}{\textit{Without}} & \textit{$\Delta$Acc} & \textit{$\Delta$R} & \textit{$\Delta$F} \\
 & & Acc & R & Acc & R & & & \\
\midrule
v4 self+calib & 5 & 85.5 & 76 & 82.5 & 68 & $+3.0$ & $+8$ & $-2$ \\
v8 full stack & 4 & 82.5 & 66 & 76.0 & 52 & $+6.5$ & $+14$ & $-1$ \\
v19 +two-sided & 1 & 87.0 & 79 & 83.0 & 70 & $+4.0$ & $+9$ & $-1$ \\
v20 +ensemble(v6) & 4 & 76.5 & 57 & 73.5 & 49 & $+3.0$ & $+8$ & $-2$ \\
v22 +dual-val(v6) & 3 & 80.0 & 63 & 75.5 & 53 & $+4.5$ & $+10$ & $-1$ \\
v23 +two-sided (freeze base) & 2 & 81.0 & 73 & 71.0 & 98 & $+10.0$ & $-25$ & $+45$ \\
v26 +case-mem & 2 & 80.0 & 86 & 77.0 & 97 & $+3.0$ & $-11$ & $+17$ \\
v31 +curriculum & 4 & 82.5 & 72 & 76.0 & 57 & $+6.5$ & $+15$ & $-2$ \\
v41 +calib-window & 3 & 77.0 & 56 & 75.0 & 54 & $+2.0$ & $+2$ & $+2$ \\
v47 +batch interaction check & 1 & 84.5 & 72 & 80.0 & 63 & $+4.5$ & $+9$ & $0$ \\
v34 +budget & 2 & 63.0 & 97 & 63.0 & 95 & $0.0$ & $+2$ & $-2$ \\
\bottomrule
\end{tabular}}
\caption{\textbf{All 11 measured self-tool ablations} (main-set final state, A/B with self-written tools enabled vs.\ disabled; R = real-image recall, F = fake-image recall). Ten of the eleven configurations lose accuracy when their own tools are disabled (mean $+4.3$ with tools enabled, up to $+10.0$). The admission-budget configuration is operating-point reversed, and its two tools leave final accuracy unchanged in either arm (real-image recall $+2$, fake-image recall $-2$; the two effects cancel).}
\label{tab:selftool_full}
\end{table}

\section{Tool Content Audit}
\label{app:audit}

All 82 distinct tool implementations admitted across the study were read and classified (multiple flags possible): 61\% emit output without a directionally parseable label (the calibration layer cannot extract evidence); 26\% return hardcoded numbers in the output string unrelated to the computation; 22\% assume the region of interest lies at the image center; 7\% compare 0--255-range pixel values against 0--1 thresholds; 7\% default to a FAKE verdict on internal exceptions (65\% of these default neutrally after contract enforcement); 6\% import forbidden libraries; 6\% exceed 200 lines (the longest, at 482 lines, enters a loop interleaved with self-generated commentary). The contract introduced after the audit (measurements only, no verdict labels; numeric output required; discrimination pre-check on labeled contrast pairs) addresses each class. Two independent implementations of ``sticker-artifact detection'' (21 lines vs.\ 46) addressed the same task with different code and different behavior; theme convergence does not imply implementation convergence, which is why tool identity, not topic, is the unit of admission. The counts reconcile with the main text: the study logged 139 tool proposals, of which 104 admission decisions covered these 82 distinct tool names (a modified tool replaces its earlier file, so the audited artifact is each name's final admitted implementation), reached through 307 sandbox repair cycles.

\section{Qualitative Artifacts}
\label{app:qualitative}

\paragraph{Experience entries (final library of the ensemble-reflection configuration v12, content verbatim).}
\begin{compactitem}
\item \textit{Context}: chest radiograph, any generator, mild Gaussian noise (gauss\_10). \textit{Cue}: diaphragm contour is unnaturally smooth and dome-shaped. \textit{Caution}: the model flags this as pathology or forgery, when it is a common generation artifact. (\textit{times-correct} 35 / \textit{times-incorrect} 5 over 40 injections; retired)
\item \textit{Context}: chest radiograph, flux.1-dev family, false alarms. \textit{Cue}: diaphragm shape and the costophrenic angle alone do not indicate forgery; these are common normal anatomical variations. \textit{Caution}: the right hemidiaphragm can show a sharp, well-defined costophrenic angle even when elevated or curved, a common finding in pediatric and young-adult patients. (\textit{times-correct} 190 / \textit{times-incorrect} 39 over 229 injections; on probation)
\item \textit{Context}: fundus, any generator, JPEG quality 25. \textit{Cue}: retinal vessels show natural, complex branching and tapering. \textit{Caution}: a ``sticker-like'' lesion appearance alone drove the model's false forgery verdicts. (\textit{times-correct} 161 / \textit{times-incorrect} 39 over 200 injections; on probation)
\item \textit{Context}: fundus, qwen family, removal edits, mild blur. \textit{Cue}: the lesion appears pasted on, without biological interaction with surrounding tissue or noise consistency. \textit{Caution}: on this family the detector misses the forgery and mis-attributes it to an unrelated shape feature. (\textit{times-correct} 127 / \textit{times-incorrect} 26 over 153 injections; on probation)
\end{compactitem}

\paragraph{Self-written tool (verbatim, 21-line sticker-artifact detector, abridged).}
\begin{tcolorbox}[colback=gray!10, colframe=gray!50, title=Abridged self-written tool (as admitted), breakable]
\footnotesize
\begin{verbatim}
def analyze(img):
    g = np.asarray(img.convert("L"), float)
    lap = np.abs(ndimage_laplacian(g))
    noise = local_std(g)
    ratio = lap / (noise + 1e-6)
    hot = (ratio > 0.5 * noise.std()).sum()
    return f"edge_to_noise=" \
           f"{ratio.mean():.3f} " \
           f"hot_pixels={hot}"
\end{verbatim}
\end{tcolorbox}
No verdict label appears in the return string; the calibration layer supplies the directional interpretation. Its admission packet showed separation of \texttt{edge\_to\_noise} between labeled real and fake probe images.

\paragraph{Discrimination table for the retired diaphragm verifier (illustrative example).}
\begin{tcolorbox}[colback=gray!10, colframe=gray!50, title=Discrimination table shown to the reflector (illustrative example)]
\footnotesize
\begin{verbatim}
tool=diaphragm_consistency (enabled r5,
      self-written, repaired r5)
   readings on MISCLASSIFIED images :
      median 0.82  IQR [0.79, 0.85]  n=11
   readings on CORRECT images :
      median 0.81  IQR [0.78, 0.84]  n=29
   separation: none (overlap 94%)
      -> NO SIGNAL
\end{verbatim}
\end{tcolorbox}
The values in this example are illustrative of the readout format; the deployed mechanism computes verdict-distribution counts of each tool's readings on the round's error cases versus correctly-judged cases from the same cluster. The reflector disabled the tool in the same round (Figure~\ref{fig:lifecycle}).

\section{Cost Accounting}
\label{app:cost}

\begin{table}[h]
\centering
\small
\setlength{\tabcolsep}{4pt}
\begin{tabularx}{\linewidth}{@{}p{0.46\linewidth}>{\raggedright\arraybackslash}X@{}}
\toprule
\textit{Quantity} & \textit{Value} \\
\midrule
Scoring per round, standard configurations & 6.4--7.2\,min (200 images; control 6.7, ensemble 7.2) \\
Scoring per round, calibration-injection variants & up to 22\,min on final-evaluation-sized batches \\
Wall-clock per 20-round configuration & 2--7 hours, depending on the calibration variant \\
Generations per learning round & $\approx$325--450, of which $>$60\% are verification calls \\
Study total & 49 configurations $\times$ 20 rounds \\
Labeled learning images, whole study & 39{,}200 (a single SFT run of the base detector uses 50{,}000) \\
\bottomrule
\end{tabularx}
\caption{\textbf{Cost accounting.} Wall-clock covers scoring and learning jointly on shared vLLM replicas; call counts cover the learning phase only (40 detector responses, 1--4 reflection sessions by configuration, 80 replay A/B calls, 200 validation A/B calls).}
\label{tab:cost}
\end{table}

Table~\ref{tab:cost} summarizes the accounting. The added mechanisms impose less than 10\% relative overhead on the standard configurations. Calibration-injection variants (the v4/v26 family) reach up to 22 minutes per round on final-evaluation-sized batches, as every query renders calibrated tool readings over the histogram history. Verification, not learning, dominates generation cost, which is an intentional design choice: replay A/B (80 calls) and validation A/B (200 calls) outnumber the 40 learning-phase responses and the reflection sessions. Counting labeled images instead of compute, the entire self-improvement phase consumes 39{,}200 labeled learning images, fewer than the 50{,}000-image labeled set of the single SFT run that produced the base detector, and performs no gradient updates.

\section{Reproducibility}
\label{app:repro}

All configurations are specified by the registry (single-key diffs on JSON configs); the harness, registry, per-round predictions for every configuration and round, every acceptance-test decision log, all experience libraries and toolboxes in final state, and the archive from the period of the verifier fault (the accidental admission-off ablation) are released. Test-set membership is fixed by seed and hash-committed; the final evaluation was built exclusively from the official held-out split with programmatic zero-overlap assertions against all training images.

\section{Use of Large Language Models}
\label{app:llm}

All reflection, tool writing, and tool repair in the self-reflection configurations is performed by the frozen detector itself (MedForge-Reasoner, served locally through vLLM); its weights and its forensic system prompt are never modified. A commercial API model, GLM-5.3-Flash, participates in exactly three of the 49 registered configurations, as the reflection model of the external experience channel (v2, v5, v33); in v5 and v33 it consequently also writes and repairs tool code. Its role is strictly that of an alternative error summarizer under the same acceptance test as every other proposal source: it receives only the round's learning-stream error digest, it never scores the fixed test sets (all scoring is deterministic code over detector outputs), it never renders acceptance-test verdicts, and it never sees the final evaluation sets. No large language model contributed to the writing, analysis, or experimentation reported in this paper beyond the roles stated here and in \S\ref{sec:setup}.

\section{Design Guidelines and Evidence}
\label{app:guidelines}

Six design guidelines for deployment-time recursive self-improvement, each traceable to a measured comparison in the main text:

\begin{compactitem}
\item \textbf{Test before admitting.} No persistent change without independent-validation acceptance; the test contributes more than any reflection mechanism ($+17.5$ versus at most $+6.5$).
\item \textbf{Calibrate tool readings against labeled history.} Convert raw readings into distributional evidence, and evaluate measurement claims on measurements (discrimination tables), never on printed text.
\item \textbf{Grant write access only on enumerable artifacts.} Experience entries, tool switches, and bounded rule lists, each individually testable and retractable. Free-form rewrites of global policy text oscillated or drifted in every configuration that allowed them.
\item \textbf{Diversify what is proposed, and leave verdict formation unchanged.} Ensemble and two-sided reflection improve the proposal supply; second-opinion arbitration and self-consistency voting change how verdicts are made, and both reduce accuracy.
\item \textbf{Protect recovery, not only stability.} Rollback that restores old snapshots discards new learning ($-4.5$); admission caps prevent repair ($-24.0$). Protection mechanisms must be able to reverse damage at least as quickly as it occurs.
\item \textbf{Replicate across seeds.} The strongest single-run result in the study did not replicate; seed-level evidence is the minimum standard for a self-improvement claim.
\end{compactitem}

\needspace{6\baselineskip}
\section{Additional Analyses}
\label{app:extra}

\paragraph{Tool selection signals can mislead.}
The second-opinion tool, which injects an external model's verdict text, showed $-19$ points counterfactual value on the validation split, yet it was re-proposed by 34 of the 49 configurations and admitted in 16, as its replay gain moved in the opposite direction from its true effect. A selection signal measured on the learning batch can point in the opposite direction from a tool's true effect at detection time. The remedy that proved effective, greedy enable/disable search scored on the validation split, applies the acceptance-test principle one level down.

% flushend balances the final page only under \flushbottom; the rest of the
% document keeps the \raggedbottom set in the preamble.
\flushbottom
\section{Failure Narratives}
\label{app:failnar}

\paragraph{Operating-point reversal, round by round.}
The external-reflector-with-toolbox configuration admitted seven changes in round 10: three experience entries, a directive rewrite, one tool enable, one tool modification, and one new tool. Each change passed both stages of the acceptance test individually (replay gains of 0, $+3$, and $+6$; validation drop of zero). Jointly they inverted the operating point: real-image recall rose to 94--100\% and fake-image recall fell to 53--72\%, and the configuration remained at the inverted operating point for the remaining ten rounds. Figure~\ref{fig:asymmetry} places this reversal within the error structure of all 47 configurations. In the reversed state the system scored near zero on the validation split in both arms of every later test, so the no-distortion condition was vacuously satisfied; the verifier health check and the batch interaction check were designed to detect and interrupt exactly this failure pattern.

\paragraph{Silent verifier failure.}
\looseness=-1 In the acceptance test's scoring helper, a generator expression unpacked \texttt{zip(samples, outs)} with its arguments swapped, so the sample variable received an output (no label field) and the output variable a sample (no prediction field); comparing two missing fields returned true for every pair; both arms compared equal at full length, every proposal passed, and the score-drop check \enlargethispage{\baselineskip} The main scoreboard compared fields explicitly and was unaffected, so the reported scores remained correct while every admission decision made during that period was invalid.
% Manual column break: balances the single-paragraph final page.
\newpage
The contradiction first appeared in one configuration's decision log (an acceptance-off arm scoring 100\% with both per-class recalls at zero), confirmed by an offline diagnostic check and manual re-implementation; the affected runs are archived as an acceptance-off ablation.

\end{document}